\documentclass[12pt]{iopart}
\usepackage[dvips]{graphicx}

\begin{document}

\title[Screening of universal Casimir terms]{``Screening'' of 
universal van der Waals -- Casimir terms by Coulomb gases in a
fully-finite two-dimensional geometry}

\author{B. Jancovici\dag\ and L. {\v S}amaj\dag\ddag}
\address{\dag\ Laboratoire de Physique Th\'eorique, Universit\'e de
Paris-Sud, B\^atiment 210, 91405 Orsay Cedex, France
(Unit\'e Mixte de Recherche no. 8627 - CNRS)}
\address{\ddag\ Institute of Physics, Slovak Academy of Sciences,
D\'ubravsk\'a cesta 9, 845 11 Bratislava, Slovakia}

\eads{\mailto{Bernard.Jancovici@th.u-psud.fr}, \mailto{fyzimaes@savba.sk}}

\begin{abstract}
This paper is a continuation of a previous one
[Jancovici B and {\v S}amaj L, 2004 J. Stat. Mech. P08006] dealing with
classical Casimir phenomena in semi-infinite wall geometries.
In that paper, using microscopic Coulomb systems, the long-ranged
Casimir force due to thermal fluctuations in conducting walls was
shown to be screened by the presence of an electrolyte between the walls
into some residual short-ranged force.
Here, we aim to extend the study of the screening (cancellation)
phenomena to universal Casimir terms appearing in the large-size expansions
of the grand potentials for microscopic Coulomb systems confined
in fully-finite 2D geometries, in particular the disc geometry.
Two cases are solved exactly: the high-temperature (Debye-H\"uckel)
limit and the Thirring free-fermion point.
Similarities and fundamental differences between fully-finite and
semi-finite geometries are pointed out.
\end{abstract}


\medskip


\maketitle

\eqnobysec

\section{Introduction}
At zero temperature, fluctuations of the quantum electromagnetic field
in vacuum manifest themselves via an attraction of two parallel
ideal-conductor plates.
This Casimir effect (for an introduction see \cite{Duplantier1})
has a universal character in the sense that it does not depend on
the material constitution of the metallic plates.
Casimir's result was extended to arbitrary temperatures and
general dielectric plates \cite{Lifshitz,Schwinger1}, and to
ideal-conductor walls of arbitrary smooth shapes \cite{Balian1}.
For a recent book and review see \cite{Mostepanko,Bordag}.
The studied models can be divided according to the geometry of
fluctuating walls into two basic sets: the semi-infinite systems,
in which at least one of the spatial coordinates is unconstrained
by the walls (e.g. two parallel plates), and the fully-finite systems
(e.g. a sphere).
The applied methods and observed Casimir phenomena usually depend
on this classification.

As concerns semi-infinite geometries, in the high-temperature limit
defined by the validity of the equipartitioning energy law, the
Casimir force becomes purely entropic \cite{Feinberg}; this force is
usually called classical since it does not depend on Planck's constant.
In the purely electrostatic models which do not incorporate the magnetic
part of the Lorentz force due to charge currents, like the system
of scalar photons \cite{Schwinger2}, the Casimir force is divided by
a factor 2.

As concerns fully-finite three-dimensional (3D) conductor systems
\cite{Balian1,Balian2}, in the high-temperature limit, the Casimir
free energy does depend on Planck's constant.
Furthermore, the presence of both electric and magnetic degrees of
freedom is necessary for obtaining a Casimir effect.
This is no longer true for classical 2D Coulomb fluids defined
in fully-finite domains.
There, the consideration of the pure Coulomb potential, defined
as the solution of the 2D Poisson equation, leads to a universal
Casimir term analogous to the one appearing in finite-size expansions 
of thermodynamic quantities for 2D critical systems with
{\em short-range} interactions among constituents.
To be more precise, it is known that, according to the principle
of conformal invariance, for a finite system of characteristic size $R$,
at a critical point, the (dimensionless) free energy has a
large-$R$ expansion of the form \cite{Blote, Affleck,Duplantier2,Cardy}
\begin{equation} \label{1.1}
\beta F = A R^2 + B R - \frac{c \chi}{6} {\rm ln}\, R
+ {\rm const} + \cdots ,
\end{equation}
where $\beta$ denotes the inverse temperature.
The coefficients $A$ and $B$ of the bulk and surface parts are
non-universal.
The coefficient of the logarithmic Casimir term is universal,
dependent only on the conformal anomaly number $c$ of the critical
theory and on the Euler number $\chi$ of the manifold on which
the system is confined.
In general, $\chi = 2-2h-b$, where $h$ is the number of handles
and $b$ the number of boundaries of the manifold
($\chi=2$ for the surface of a sphere, $\chi=1$ for a disc,
$\chi=0$ for an annulus or a torus).
A simple example is the Gaussian model \cite{Houches} which is
critical at all temperatures, with the conformal anomaly number
$c=1$.
At any temperature of the conducting regime, the grand potential
of a classical 2D Coulomb system of characteristic size $R$ is supposed
to exhibit a large-$R$ expansion of type (\ref{1.1}), with however a +
sign in front of the logarithmic term:
\begin{equation} \label{1.2}
\beta \Omega_{\rm Coulomb} = A R^2 + B R + \frac{\chi}{6} {\rm ln}\, R
+ {\rm const} + \cdots .
\end{equation}
Plausible arguments for a critical-like behaviour were first given
for Coulomb gases with periodic boundary conditions \cite{Forrester},
then for Coulomb systems confined to a domain by (vacuum) plain hard
walls \cite{Jancovici1}, by inert ideal-conductor walls 
(i.e. when the electric potential obeys Dirichlet boundary conditions) 
\cite{Jancovici2} and finally by ideal-dielectric boundaries 
(i.e. when the electric potential obeys Neumann boundary conditions) 
\cite{Tellez}.
The explicit checks were done in the Debye-H\"uckel limit
\cite{Torres}, at the free-fermion point of the Thirring
representation of the symmetric two-component plasma, based
on the formalism developed in \cite{Cornu}, and for the one-component
plasma at $\beta=2$ in a disc \cite{Jancovici1,Jancovici2}.
Only recently, a direct derivation of the universal finite-size
correction term was done, in the whole stability range of temperatures,
for the specific cases of the symmetric two-component plasma
living on the surface of a sphere \cite{Jancovici3} and in a disc
surrounded by vacuum \cite{Samaj1}.
In both cases, the universal prefactor to the ${\rm ln}\, R$
correction term in (\ref{1.2}) was related to the bulk second moment
of the density structure factor which is known \cite{Janco}.
This is very different from the semi-infinite geometries where
the universality of the Casimir force is related to the second moment
sum rule for the {\em charge} structure factor \cite{Jancovici2,Buenzli}.

It is generally believed that, for semi-infinite systems, the presence
of an electrolyte between fluctuating conductor walls screens the
long-ranged Casimir force to some residual short-ranged force
\cite{Attard,Dean}.
The study of Casimir phenomena via fully microscopic Coulomb models
has the advantage of a coherent description of electrostatic fluctuations
inside conducting walls and the image forces acting on the electrolyte
particles, without any ad hoc separation Ansatz used in usual
macroscopic treatments \cite{Attard}.
We used this strategy to show the screening effect of Casimir forces for
semi-infinite geometries in paper \cite{Jancovici4}, in what follows
referred to as I.

The aim of the present work is to extend the study of the screening
(cancellation) phenomena to universal Casimir terms appearing
in the large-size expansions of the grand potential (\ref{1.2})
for the Coulomb systems confined in fully-finite 2D geometries,
in particular the disc geometry.
The confining disc walls as well as the electrolyte inside the disc
are modelled by two different microscopic two-component plasmas of
point-like particles with $+/-$ unit charges in thermal equilibrium.
Two cases are solved exactly: the high-temperature (Debye-H\"uckel)
limit $\beta\to 0$ and the Thirring free-fermion point $\beta=2$
corresponding to the collapse of positive and negative pairs
of point-like charges.
From the technical point of view, the circular symmetry of the problem
leads to infinite summations over specific products of modified
Bessel functions which have to be evaluated by using the asymptotic
Debye expansion; we apply a few technicalities which help us to
simplify a relatively complicated algebra.
Similarities and differences with respect to screening phenomena
in semi-infinite geometries, described in paper I, are pointed out.

\begin{figure}[h]
\begin{center}
\includegraphics{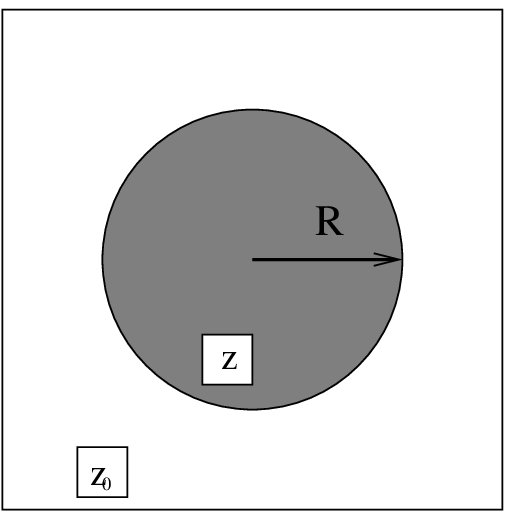}
\mbox{}
\mbox{Figure 1. Disc geometry.}
\end{center}
\end{figure}

The model is defined as follows.
We shall consider Coulomb-gas systems of point-like particles with
symmetric $+/-$ unit charges.
Thermal equilibrium is treated in the grand canonical ensemble
characterized by the inverse temperature $\beta$ and by the couple
of equivalent (there is no external electrostatic potential),
possibly position-dependent, particle fugacities
$z_+({\bf r}) = z_-({\bf r}) = z({\bf r})$.
In the disc geometry presented in figure 1, there are two domains:
the disc of radius $R$, $\Lambda_R = \{ {\bf r}, \vert {\bf r}\vert < R \}$,
and its complement ${\bar \Lambda}_R = \{ {\bf r},\vert{\bf r}\vert\ge R \}$.
The electrolyte in $\Lambda_R$ is modelled by a two-component
plasma with the particle fugacity denoted by $z$.
The wall ${\bar \Lambda}_R$ is modelled by another two-component plasma
with the particle fugacity $z_0$; the choice $z_0> 0$ corresponds to
a conducting wall $(\epsilon\to\infty)$ while $z_0=0$ corresponds to
vacuum (plain hard wall with $\epsilon=1$).
In 2D, the Coulomb potential $v$ at a spatial position ${\bf r}$,
induced by a unit charge at the origin ${\bf 0}$, is the solution
of the 2D Poisson equation
\begin{equation} \label{1.3}
\Delta v({\bf r}) = - 2 \pi \delta({\bf r}) .
\end{equation}
Explicitly,
\begin{equation} \label{1.4}
v({\bf r}) = -{\rm ln}\,( \vert {\bf r}\vert /a)
\end{equation}
where $a$ is a free length scale, which only determines the zero of
the potential and should not enter statistical mean values.
The interaction energy of charged particles $\{ i, q_i \}$,
immersed in a homogeneous medium of dielectric constant = 1, is
$\sum_{i<j} q_i q_j v(\vert {\bf r}_i-{\bf r}_j \vert)$.

The paper is organized as follows.

In general section 2, the Coulomb gas confined to some 2D regions
is shown to be equivalent, in the ideal-conductor limit, to the
massless Gaussian model defined in the complementary empty regions.
Since in the derivation \cite{Houches} of the large-size expansion
of the free energy (\ref{1.1}) for the critical Gaussian model with
$c=1$ only the curvature of the boundary is used, this explains
the difference in sign of the logarithmic term between
(\ref{1.1}) and (\ref{1.2}).

In section 3, the underlying 2D fully-finite Coulomb system is solved
in the Debye-H\"uckel limit.
All possible realizations of the model are considered, and the
cancellation of the universal Casimir terms is documented when
both the disc domain and its complement are occupied by a Coulomb gas.
Fundamental differences between fully-finite and semi-infinite
geometries are pointed out.

The exact solution of the model system at the Thirring free-fermion
point $\beta=2$ is presented in section 4.
It is shown that basic features of Casimir phenomena, predicted by
the mean-field theory, persist also at this specific temperature.

A brief recapitulation is given in section 5.

\section{The Gaussian model approach}
The universal term $(\chi/6)\ln R$ in the grand potential
(\ref{1.2}) of a finite 2D Coulomb system closely resembles
the logarithmic term in (\ref{1.1}) valid for critical systems.
At first sight, it is surprising that a Coulomb system,
with short-range particle correlations, exhibits a critical-like behaviour.
It has been argued \cite{Jancovici1,Jancovici2} that this is related to
the existence of long-ranged critical-like correlations of
the electric potential; reasons for the difference in sign of the
logarithmic term between (\ref{1.1}) and (\ref{1.2}) have been given.
A slightly different, and we believe more convincing, argument
will now be presented.

\subsection{General formalism}
In the present subsection, we consider a 2D manifold on which some
regions $C$ are classical ideal conductors and some regions $E$ are empty.
The ideal conductors are obtained as the infinite fugacity limit
of conductors having some microscopic structure, which amounts to take
the limit in which the microscopic length scale goes to zero.
Thus, the results which will be obtained in this limit are expected
to be valid when the length scales under consideration are much larger
than the microscopic lengths.
For simplicity, we restrict ourselves to an infinite-plane manifold,
$C\cup E = R^2$, and start with conductors which are symmetric Coulomb 
gases of the kind described in the Introduction, but some generalizations 
are straightforward.

The interaction energy $E_N$ of $\pm 1$ charged particles
$\{ q_j, {\bf r}_j \}_{j=1}^N$ is expressible in terms of the
microscopic charge density 
${\hat \rho}_N({\bf r}) = \sum_{j=1}^N q_j \delta({\bf r}-{\bf r}_j)$ 
as follows
\begin{equation} \label{2.1}
E_N(\{ q_j,{\bf r}_j \}) = \frac{1}{2} \int_{R^2} {\rm d}^2 r 
\int_{R^2} {\rm d}^2 r' {\hat\rho}_N({\bf r}) v(\vert {\bf r}-{\bf r}'\vert)
{\hat\rho}_N({\bf r}') - \frac{1}{2} N v(0) ,
\end{equation}
where $v(0)$ is the (diverging) self-energy.
The thermodynamic properties of the Coulomb system are determined
by the grand partition function $\Xi$ defined as follows
\numparts
\begin{equation} \label{2.2a}
\Xi = \sum_{N_+=0}^{\infty} \sum_{N_-=0}^{\infty}
\frac{z_+^{N_+}}{N_+!} \frac{z_-^{N_-}}{N_-!} Q(N_+,N_-) , 
\end{equation}
where
\begin{equation} \label{2.2b}
Q(N_+,N_-) = \int_C \prod_{j=1}^N {\rm d}^2 r_j 
\exp\left[ - \beta E_N(\{ q_j, {\bf r}_j \}) \right]
\end{equation}
\endnumparts
is the configuration integral of $N_+$ positive and $N_-$ negative
charges, $N=N_+ + N_-$ and $z_+ = z_- = z$ are the equivalent
fugacities of charged particles.
Let us insert the energy representation (\ref{2.1}) into (\ref{2.2b}). 
The self-energy simply renormalizes the particle fugacities.
With regard to the fact that, according to (\ref{1.3}), 
$-\Delta/(2\pi)$ is the inverse operator of the Coulomb potential 
$v({\bf r})$, the standard Hubbard-Stratonovich transformation
(see e.g. ref. \cite{Samuel}) can be used to express the integral
bilinear term as
\begin{eqnarray} 
\fl \exp\left[ - \frac{\beta}{2} \int_{R^2} {\rm d}^2 r \int_{R^2} 
{\rm d}^2 r' {\hat\rho}_N({\bf r}) v(\vert {\bf r}-{\bf r}'\vert)
{\hat\rho}_N({\bf r}') \right] 
\nonumber \\
= \frac{\int {\cal D}\phi \exp\left[ \int_{R^2} {\rm d}^2 r
\left( \frac{\beta}{4\pi} \phi \Delta \phi + {\rm i} \beta \phi {\hat\rho}_N 
\right) \right]}{\int {\cal D}\phi \exp\left( \int_{R^2} {\rm d}^2 r 
\frac{\beta}{4\pi} \phi \Delta \phi \right)} , \label{2.3} 
\end{eqnarray}
where $\phi({\bf r})$ is a real scalar field and $\int {\cal D} \phi$
denotes the functional integration over this field.
The consequent factorization of the contributions from $(N_+,N_-)$
particle states in (\ref{2.2a}) then allows to express 
the grand partition function of the system in the form \cite{Minnhagen}  
\numparts
\begin{equation} \label{2.4a}
\Xi = \frac{\int{\mathcal D}\phi\exp(-S[z({\bf r})])}
{\int{\mathcal D}\phi\exp(-S[0])},
\end{equation}
with
\begin{equation} \label{2.4b}
S[z({\bf r})]=\int_{C\cup E} {\rm d}^2 r\left[\frac{\beta}{4\pi}\phi({\bf r})
(-\Delta)\phi({\bf r})
-2z({\bf r})\cos(\beta\phi({\bf r}))\right].
\end{equation}
\endnumparts
being the 2D Euclidean action of the sine-Gordon theory and
\begin{equation} \label{2.5}
z({\bf r}) = \left\{
\begin{array}{ll}
z & \mbox{for ${\bf r} \in C$,} \cr
0 & \mbox{for ${\bf r} \in E$.}
\end{array}  \right.
\end{equation} 
The normalization of $z$ is fixed by the short-distance expansion 
of the two-point correlation function
\begin{equation} \label{2.6}
\langle {\rm e}^{{\rm i}\beta \phi({\bf r})} 
{\rm e}^{-{\rm i}\beta \phi({\bf r}')} \rangle \sim
\vert {\bf r}-{\bf r}' \vert^{-\beta} \qquad 
\mbox{as $\vert {\bf r}-{\bf r}' \vert \to 0$}
\end{equation}
under which the self-energy factor disappears from statistical
relations; for more details see ref. \cite{Samaj2} and the references
cited therein.

Since the scalar $\phi$-field has to be regular at infinity,
the term $\phi (-\Delta) \phi$ in (\ref{2.4b}) can be transformed
via the integration by parts into $\vert \nabla \phi \vert^2$.
The sine-Gordon action (\ref{2.4b}) thus takes its minimum
at a $\phi({\bf r})$ constant in space.
Due to a discrete symmetry $\phi\to \phi + 2\pi n/\beta$
with any integer $n$, the action has infinitely many ground states
$\vert 0_n \rangle$ characterized by the associate expectation
values of the field $\langle \phi \rangle_n = 2 \pi n / \beta$.
These ground states become all degenerate when the size of the Coulomb
domain $\vert C\vert$ is large \cite{Zamolodchikov}.
It is therefore sufficient to develop the sine-Gordon action
(\ref{2.4b}), on the classical level as well as on higher
quantum-correction levels, around any one of these ground states
\cite{Zamolodchikov},
say $\vert 0_0 \rangle$ with $\langle \phi \rangle_0 = 0$.
Now, in the ideal-conductor limit $z\to\infty$, 
the minimum of $-2z\cos(\beta\phi)$ around $\phi=0$ becomes infinitely
sharp and $\phi({\bf r})$ identically vanishes in the regions $C$.
Up to an irrelevant multiplicative constant,
the numerator of (\ref{2.4a}) becomes
\begin{equation} \label{2.7}
Z_{\rm G}=\int{\mathcal D}\phi\exp\left[-\frac{\beta}{4\pi}
\int_E{\rm d}^2 r\phi({\bf r})(-\Delta)\phi({\bf r})\right],
\end{equation}
i.e. the partition function of the massless Gaussian model in the empty
region(s).
The boundary condition for $\phi$ is that it vanishes at the interface(s)
of $E$ and $C$, since it has to be continuous and vanishes in $C$.
The denominator of (\ref{2.4a}) is just an (infinite) constant,
independent of the geometry of the regions $E$ and $C$.

Although $\phi$ has some resemblance with the electric potential, it
also has some drastically different properties.
For instance, in a bulk Coulomb gas, the $\phi$ of the sine-Gordon
representation has short-range two-point correlations while
the electric potential has long-range ones \cite{Caillol}.
Here, the electric potential has fluctuations in the conductor(s)
(these fluctuations survive in the ideal-conductor limit) and
by continuity the electric potential has fluctuations at the interface(s)
of $E$ and $C$, while $\phi$ is identically zero at the interface(s).

\subsection{Parallel conducting plates in $\nu$ dimensions}
In the case of two parallel conducting plates, at a distance $d$ of each
other, separated by vacuum, the Gaussian model (\ref{2.7}), generalized
to $\nu$ dimensions, has been shown \cite{Blote,Jancovici2} to have in its
free energy per unit area of one plate a universal $d$-dependent term $F$
such that the corresponding electrostatic Casimir force per unit area
$-\partial F/\partial d$ is attractive and given by
\begin{equation}  \label{2.8}
-\beta\frac{\partial F}{\partial d}=
-\frac{s_{\nu-1}\Gamma(\nu)\zeta(\nu)}{(2\pi)^{\nu-1}(2d)^{\nu}},
\end{equation}
where $s_{\nu}$ is the area of the unit sphere in $\nu$ dimensions and
where $\Gamma$ and $\zeta$ are the gamma function and the Riemann zeta
function, respectively.
In the Appendix, we give another derivation of (\ref{2.8}) for
the Gaussian model.
In subsection 2.2. of I, we have obtained the same Casimir force (\ref{2.8})
by using the fluctuations of the electric potential
(which do not vanish at the surface of the plates).
This is a check about the validity of the present method using
the Gaussian model with simple Dirichlet boundary conditions
for the $\phi$ field.

\subsection{A hole in a 2D ideal conductor}
We now come back to two dimensions.
We consider an ideal conductor filling the whole plane except for
an empty hole of characteristic size $R$, with a smooth boundary.
It has been shown \cite{Houches} that the Gaussian model has
a free energy of the form (\ref{1.1}) with $c=1$ and $\chi=1$.
Thus, $\beta$ times the grand potential of the system has the
universal term $-(1/6)\ln R$.
If the hole is a disc in a plane, $R$ may be chosen as its radius.

\subsection{A 2D Coulomb gas surrounded by an ideal conductor}
We now consider the opposite geometry of an ideal conductor filling some
connected region $C$ of the plane, with a smooth boundary, surrounded by
vacuum. Now the relevant Gaussian model fills the exterior of $C$.
In the derivation of ref. \cite{Houches}, the curvature of the boundary
is used, and it is easy to see that the present geometry differs from
the one of Section 2.3 by a change of sign of the curvature, leading to
the universal term $+(1/6)\ln R$ in (\ref{1.2}).

\section{Debye-H\"uckel theory}

\subsection{General result}
The high-temperature limit of the model system in figure 1 is described
by the Debye-H\"uckel theory.
For the general formalism, see e.g. section 3.1. of paper I.
The formalism applied to the 2D symmetric Coulomb gas
can be briefly summarized as follows.

In the bulk (homogeneous) regime, the total particle number density $n$
as a function of the fugacity $z$ is given by
\begin{equation} \label{3.1}
\frac{n^{1-(\beta/4)}}{z a^{\beta/2}} =
2 \left( \frac{\pi\beta}{2} \right)^{\beta/4}
\exp \left( \frac{\beta C}{2} \right) ,
\end{equation}
where $C$ is the Euler's constant.
The inverse Debye length is defined by $\kappa^2 = 2\pi \beta n$.

In the inhomogeneous regime, the whole system domain $\Lambda$
can be separated into disjunct physically non-equivalent subdomains,
$\Lambda = \cup_{\alpha} \Lambda^{(\alpha)}$.
Within the grand canonical formalism, each subdomain is characterized
by a constant fugacity, $z({\bf r}) = z^{(\alpha)}$  for
${\bf r}\in \Lambda^{(\alpha)}$; the choice $z^{(\alpha)}=0$ corresponds
to a vacuum subdomain with no particles allowed to occupy the space.
The corresponding ``bulk'' particle density $n^{(\alpha)}$ is related to
the particle fugacity $z^{(\alpha)}$ via (\ref{3.1}), and the corresponding
inverse Debye length is $\kappa_{\alpha} = (2\pi\beta n^{(\alpha)})^{1/2}$.
One introduces the screened Coulomb potential $G$ which obeys
within each domain $\Lambda^{(\alpha)}$ the differential equation
\begin{equation} \label{3.2}
\left[ \Delta_1 - \kappa_{\alpha}^2 \right] G({\bf r}_1,{\bf r}_2)
= - 2\pi \delta( {\bf r}_1 - {\bf r}_2 ), \qquad
{\bf r}_1 \in \Lambda^{(\alpha)} .
\end{equation}
Here, the spatial position of the source point ${\bf r}_2$ is arbitrary.
These equations are supplemented by the usual electrostatic conditions
at each subdomain boundary $\partial \Lambda^{(\alpha)}$: $G$ and its
normal derivative with respect to the boundary surface,
$\partial_{\perp} G$, are continuous at $\partial\Lambda^{(\alpha)}$.
The leading $\beta$-correction to the constant particle density
$n^{(\alpha)}$ in the subdomain $\Lambda^{(\alpha)}$ is then determined
by linearizing the exponential in the expression
\begin{equation} \label{3.3}
n^{(\alpha)}({\bf r}) = 2 z^{(\alpha)} \exp \left\{ \frac{\beta}{2}
\lim_{{\bf r}'\to {\bf r}}
\left[ - G({\bf r},{\bf r}') + v(\vert {\bf r} - {\bf r}' \vert) \right] 
\right\} 
\end{equation}
where the 2D Coulomb potential $v$ is defined in equation (\ref{1.4}).

We now apply the above formalism to the geometry of interest
presented in figure 1.
The inverse Debye length will be denoted by $\kappa$ for the
disc domain $\Lambda_R = \{ {\bf r}, r<R \}$ and by $\kappa_0$
for the complement wall domain ${\bar\Lambda}_R = \{ {\bf r}, r>R \}$.
We first consider the case when both $\kappa$ and $\kappa_0$
are nonzero.
Let the source point ${\bf r}_2$ be first in the disc domain,
i.e. $r_2<R$.
Equations (\ref{3.2}) then take the form
\numparts
\begin{eqnarray}
\left[ \Delta_1 - \kappa^2 \right] G({\bf r}_1,{\bf r}_2)
= - 2\pi \delta({\bf r}_1-{\bf r}_2) , \qquad & (r_1 < R) , \label{3.4a} \\
\left[ \Delta_1 - \kappa_0^2 \right] G({\bf r}_1,{\bf r}_2)
= 0 , & (r_1 > R). \label{3.4b}
\end{eqnarray}
\endnumparts
In terms of polar coordinates, the solution of these equations can be
written as an expansion of the form
\numparts
\begin{eqnarray}
\fl G({\bf r}_1,{\bf r}_2) = \sum_{l=-\infty}^{\infty}
\left[ I_l(\kappa r_<) K_l(\kappa r_>) + a_l I_l(\kappa r_1) I_l(\kappa r_2)
\right] \exp[{\rm i}l(\varphi_1-\varphi_2)], \quad
& ( r_1 < R), \label{3.5a} \\
\fl G({\bf r}_1,{\bf r}_2) = \sum_{l=-\infty}^{\infty}
b_l K_l(\kappa_0 r_1) I_l(\kappa r_2) \exp[{\rm i}l(\varphi_1-\varphi_2)],
& ( r_1 > R). \label{3.5b}
\end{eqnarray}
\endnumparts
Here, $r_< = {\rm min}\{ r_1,r_2 \}$ and $r_> = {\rm max}\{ r_1,r_2 \}$.
$I_l$ and $K_l$ are modified Bessel functions possessing the symmetry
$I_l = I_{-l}$, $K_l = K_{-l}$.
They satisfy the same differential equation
\begin{equation} \label{3.6}
f'' + \frac{1}{x} f' - \left( 1 + \frac{l^2}{x^2} \right) f = 0 ,
\qquad \mbox{$f=I_l(x)$ or $K_l(x)$,}
\end{equation}
but exhibit different asymptotic behaviours \cite{Gradshteyn}:
\begin{equation} \label{3.7}
I_l(x) { \atop \stackrel{\sim}{\scriptstyle{x\to\infty}} }
\frac{\rme^x}{\sqrt{2\pi x}} , \qquad
K_l(x) { \atop \stackrel{\sim}{\scriptstyle{x\to\infty}} }
\left( \frac{\pi}{2 x} \right)^{1/2} \rme^{-x}
\end{equation}
and
\begin{equation} \label{3.8}
I_l(x) { \atop \stackrel{\sim}{\scriptstyle{x\to 0}} }
\frac{1}{\vert l\vert !} \left( \frac{x}{2} \right)^{\vert l \vert},
\qquad K_l(x) { \atop \stackrel{\sim}{\scriptstyle{x\to 0}} }
\frac{\vert l\vert !}{2\vert l\vert}
\left( \frac{x}{2} \right)^{- \vert l \vert},
\end{equation}
except for the special $l=0$ case of $K_0(x) \sim - \ln(x/2)-C$
in the limit $x\to 0$ where $C$ is Euler's constant.
We see that, in representations (\ref{3.5a}) and (\ref{3.5b}) of $G$,
these asymptotic behaviours ensure the regularity of $G$ at
the origin and at infinity.
The coefficients $a_l$ and $b_l$ are determined by the above defined
boundary conditions for $G({\bf r}_1,{\bf r}_2)$ at $r_1 = R$:
\numparts
\begin{eqnarray}
K_l(\kappa R) + a_l I_l(\kappa R) = b_l K_l(\kappa_0 R) , \label{3.9a} \\
\kappa \left[ K_l'(\kappa R) + a_l I_l'(\kappa R) \right]
= \kappa_0 b_l K_l'(\kappa_0 R) . \label{3.9b}
\end{eqnarray}
\endnumparts
Using the recursion formulas for the modified Bessel functions
\numparts
\begin{eqnarray}
x I_l'(x) = x I_{l\pm 1}(x) \pm l I_l(x) , \label{3.10a} \\
x K_l'(x) = - x K_{l\pm 1}(x) \pm l K_l(x) , \label{3.10b}
\end{eqnarray}
\endnumparts
and the Wronskian relation \cite{Gradshteyn}
\begin{equation} \label{3.11}
I_l(x) K_{l+1}(x) + I_{l+1}(x) K_l(x) = \frac{1}{x} ,
\end{equation}
equations (\ref{3.9a}) and (\ref{3.9b}) give
\numparts
\begin{eqnarray}
a_l = - \frac{K_l(\kappa R)}{I_l(\kappa R)} + \frac{1}{R W_l}
\frac{K_l(\kappa_0 R)}{I_l(\kappa R)} , \label{3.12a} \\
b_l = \frac{1}{R W_l} , \label{3.12b}
\end{eqnarray}
\endnumparts
where the auxiliary quantity $W_l$ is given by
\begin{eqnarray}
\fl W_l = \kappa I_l'(\kappa R) K_l(\kappa_0 R) -
I_l(\kappa R) \kappa_0 K_l'(\kappa_0 R) \nonumber \\
\lo= \kappa I_{l+1}(\kappa R) K_l(\kappa_0 R) +
I_l(\kappa R) \kappa_0 K_{l+1}(\kappa_0 R) \label{3.13} \\
\lo= \kappa I_{l-1}(\kappa R) K_l(\kappa_0 R) +
I_l(\kappa R) \kappa_0 K_{l-1}(\kappa_0 R) . \nonumber
\end{eqnarray}
Note the symmetries $a_l = a_{-l}$, $b_l = b_{-l}$ and $W_l = W_{-l}$.
With regard to the differential equation (\ref{3.6}) obeyed by
the modified Bessel functions, it is easy to show that
$W_l$ fulfills the equality
\begin{equation} \label{3.14}
\frac{\partial}{\partial R} {\rm ln}\, ( R W_l ) = \frac{1}{W_l}
(\kappa^2 - \kappa_0^2) I_l(\kappa R) K_l(\kappa_0 R) .
\end{equation}
The same procedure can be applied when the source point ${\bf r}_2$
lies outside of the disc, i.e. $r_2>R$.
For $r_1>R$, one obtains the solution of the form
\begin{equation} \label{3.15}
\fl G({\bf r}_1,{\bf r}_2) = \sum_{l=-\infty}^{\infty}
\left[ I_l(\kappa_0 r_<) K_l(\kappa_0 r_>)
+ c_l K_l(\kappa_0 r_1) K_l(\kappa_0 r_2) \right]
\exp[{\rm i}l(\varphi_1-\varphi_2)] .
\end{equation}
The coefficients $c_l$ are determined by the boundary conditions for
$G$ at $r_1=R$ as follows
\begin{equation} \label{3.16}
c_l = - \frac{I_l(\kappa_0 R)}{K_l(\kappa_0 R)} +
\frac{1}{R W_l} \frac{I_l(\kappa R)}{K_l(\kappa_0 R)} .
\end{equation}
They fulfill the symmetry $c_l = c_{-l}$.

To obtain the density profile, note that in relations (\ref{3.5a}) and
(\ref{3.15}) the first terms in the sums over $l$ correspond to
the expansion of the modified Bessel functions
$K_0(\kappa\vert {\bf r}_1-{\bf r}_2\vert)$ and
$K_0(\kappa_0\vert {\bf r}_1-{\bf r}_2\vert)$, respectively.
In formula (\ref{3.3}) for the particle density, these bulk
contributions, minus the pure Coulomb potential $v$, imply the
density-fugacity relationship (\ref{3.1}).
The second terms in the sums over $l$ in equations (\ref{3.5a}) and
(\ref{3.15}) are ``reflected'' contributions due to the boundary
at $r=R$.
After the linearization of (\ref{3.3}) in $\beta$, they lead to
\numparts
\begin{eqnarray}
n(r) = n - \frac{\beta n}{2} \sum_{l=0}^{\infty} \mu_l a_l
I_l^2(\kappa r) , & (r < R) , \label{3.17a} \\
n(r) = n_0 - \frac{\beta n_0}{2} \sum_{l=0}^{\infty} \mu_l c_l
K_l^2(\kappa_0 r) , \qquad & (r > R) . \label{3.17b}
\end{eqnarray}
\endnumparts
Here, $\mu_l$ is the Neumann factor: $\mu_0=1$ and $\mu_l=2$ for $l\ge 1$.

Having at one's disposal the density profile, the grand canonical
partition function $\Xi_R(\kappa,\kappa_0)$ can be deduced in
the following way.
For the present geometry, $\Xi_R$ is defined by
\begin{equation} \label{3.18}
\fl \Xi_R = \sum_{N_+,N_-=0}^{\infty} \frac{1}{N_+! N_-!} \prod_{i=1}^N
\left( \int_0^R \rmd^2 r_i \, z + \int_R^{\infty} \rmd^2 r_i \, z_0 \right)
\exp\left[ - \beta \sum_{i<j} q_i q_j
v(\vert {\bf r}_i - {\bf r}_j \vert) \right] ,
\end{equation}
where $N_+$ $(N_-)$ is the number of positively (negatively)
charged particles and $N = N_+ + N_-$.
The averaged particle density at position ${\bf r}$ is given by
$n({\bf r}) = \langle \sum_i \delta({\bf r}-{\bf r}_i)\rangle$.
Combining this with equation (\ref{3.18}), one finds that
\begin{equation} \label{3.19}
\frac{\partial}{\partial R} {\rm ln}\, \Xi_R =
2 \pi R \left[ n(R^-) - n(R^+) \right] ,
\end{equation}
where $R^-$ $(R^+)$ means the left (right) limit to $R$.
With respect to the density profile [equations (\ref{3.17a}) and
(\ref{3.17b}) with coefficients $a_l$ and $c_l$ given by (\ref{3.12a})
and (\ref{3.16}), respectively], this relation takes the explicit form
\begin{eqnarray}
\fl \frac{\partial}{\partial R} {\rm ln}\, \Xi_R = 2 \pi R ( n - n_0 )
- \frac{1}{2} (\kappa^2 - \kappa_0^2) \sum_{l=0}^{\infty}
\mu_l \frac{1}{W_l} I_l(\kappa R) K_l(\kappa_0 R) \nonumber \\
+ \frac{R}{2} \sum_{l=0}^{\infty} \mu_l
\left[ \kappa^2 I_l(\kappa R) K_l(\kappa R) -
\kappa_0^2 I_l(\kappa_0 R) K_l(\kappa_0 R) \right] . \label{3.20}
\end{eqnarray}
The integration with respect to $R$ of the second term on the rhs
of (\ref{3.20}) can be done by using (\ref{3.14}) and the integration
of the last term follows from the indefinite integration formula
for Bessel functions \cite{Erdelyi}
\begin{eqnarray}
\fl \int \rmd x \, x I_l(x) K_l(x) = \frac{x^2}{4} \left[
2 I_l(x) K_l(x) + I_{l+1}(x) K_{l-1}(x) + I_{l-1}(x) K_{l+1}(x) \right]
\nonumber \\
\lo= \frac{1}{2}
\left[ (x^2+l^2) I_l(x) K_l(x) - x^2 I_l'(x) K_l'(x) \right] . \label{3.21}
\end{eqnarray}
Thence, from (\ref{3.20}) one gets
\begin{eqnarray}
\fl {\rm ln}\, \Xi_R(n,n_0) = {\rm const} + \pi R^2 ( n - n_0 )
- \frac{1}{2} \sum_{l=0}^{\infty} \mu_l {\rm ln}\, ( R W_l ) \nonumber \\
+ \frac{1}{4} \sum_{l=0}^{\infty} \mu_l \left\{
\left[ (\kappa R)^2 +l^2 \right] I_l(\kappa R) K_l(\kappa R)
-(\kappa R)^2 I_l'(\kappa R) K_l'(\kappa R) \right\}
\label{3.22} \\
- \frac{1}{4} \sum_{l=0}^{\infty} \mu_l \left\{
\left[ (\kappa_0 R)^2 +l^2 \right] I_l(\kappa_0 R) K_l(\kappa_0 R)
-(\kappa_0 R)^2 I_l'(\kappa_0 R) K_l'(\kappa_0 R) \right\} .
\nonumber
\end{eqnarray}
The integration constant is fixed by considering the $R\to 0$ limit.
With the aid of the asymptotic formulas (\ref{3.8}) one gets,
for instance, $\lim_{R\to 0} ( R W_l ) = (\kappa/\kappa_0)^{\vert l\vert}$.
After simple algebra, the final result reads
\begin{eqnarray}
\fl {\rm ln}\, \Xi_R(n,n_0) = {\rm ln}\, \Xi_{R=0} + \pi R^2 ( n - n_0 )
- \frac{1}{2} \sum_{l=0}^{\infty} \mu_l {\rm ln}\, \left[ R W_l \left(
\frac{\kappa_0}{\kappa} \right)^l \right] \nonumber \\
+ \frac{1}{4} \sum_{l=0}^{\infty} \mu_l \left\{
\left[ (\kappa R)^2 +l^2 \right] I_l(\kappa R) K_l(\kappa R)
-(\kappa R)^2 I_l'(\kappa R) K_l'(\kappa R) - l \right\}
\label{3.23} \\
- \frac{1}{4} \sum_{l=0}^{\infty} \mu_l \left\{
\left[ (\kappa_0 R)^2 +l^2 \right] I_l(\kappa_0 R) K_l(\kappa_0 R)
-(\kappa_0 R)^2 I_l'(\kappa_0 R) K_l'(\kappa_0 R) - l \right\} .
\nonumber
\end{eqnarray}
Here, $\Xi_{R=0}$ is the grand canonical partition function of the system
with zero disc radius, given by the obvious relation
$\lim_{\Lambda\to\infty} {\rm ln}\, \Xi_{R=0}/\vert\Lambda\vert =
\beta p(n_0) = n_0 [1 - (\beta/4)]$.
Formula (\ref{3.23}) was derived under the assumption that both
particle densities $n$ and $n_0$ are nonzero.

The limit $n\to 0$ corresponds to no particles present inside the disc,
i.e. the vacuum disc hole surrounded by the fluctuating conductor wall.
Taking the $n\to 0$ limit in (\ref{3.23}), one gets
\begin{eqnarray}
\fl {\rm ln}\, \Xi_R(n=0,n_0) = {\rm ln}\, \Xi_{R=0} -\pi R^2 n_0
- \frac{1}{2} \sum_{l=0}^{\infty} \mu_l
{\rm ln} \left[ 2 \left( \frac{\kappa_0 R}{2} \right)^{l+1}
\frac{1}{l!} K_{l+1}(\kappa_0 R) \right] \label{3.24} \\
- \frac{1}{4} \sum_{l=0}^{\infty} \mu_l \left\{
\left[ (\kappa_0 R)^2 +l^2 \right] I_l(\kappa_0 R) K_l(\kappa_0 R)
-(\kappa_0 R)^2 I_l'(\kappa_0 R) K_l'(\kappa_0 R) - l \right\} .
\nonumber
\end{eqnarray}

The limit $n_0\to 0$ corresponds to no particles present outside
of the disc, i.e. the Coulomb system in the disc surrounded
by vacuum (the plain hard wall).
One can take the $n_0\to 0$ limit in every term of equation (\ref{3.23}),
except for the $l=0$ term in the first summation on the rhs of
that equation.
This $l=0$ term makes a problem due to the logarithmic divergence of
$K_0(\kappa_0 R)$ in $W_0$ in the limit $\kappa_0\to 0$.
The problem with the $l=0$ term, caused by the fact that the effective
Coulomb potential $G(r)$ is not screened at asymptotically large distances
$r$ in the present limit $\kappa_0\to 0$, was discussed e.g.
in \cite{Jancovici5}).
According to this reference, when calculating the particle density
(\ref{3.17a}) one finds that the coefficient $a_0$ depends on the
free length scale $a$ of the 2D Coulomb potential (\ref{1.4}).
Since a statistical mean value should not depend on $a$,
the limit $a\to\infty$, which puts the zero of the Coulomb potential
to infinity, has to be considered.
Our formalism in the limit $\kappa_0\to 0$ is equivalent to that
outlined in \cite{Jancovici5} if one sets $\kappa_0 = 1/a$.
Thus, returning to the particle density (\ref{3.17a}), the auxiliary
quantity $W_l$ (\ref{3.13}), considered in the limit $\kappa_0\to 0$
and for $R>0$, behaves like
\begin{equation} \label{3.25}
W_l \sim \kappa I_{l-1}(\kappa R) K_l(\kappa_0 R)
\qquad \mbox{for $l\ge 0$,}
\end{equation}
and the coefficients $a_l$, given by (\ref{3.12a}), take the form
\begin{equation} \label{3.26}
a_l = \frac{K_{l-1}(\kappa R)}{I_{l-1}(\kappa R)} ,
\qquad l=0,1, \ldots .
\end{equation}
On the basis of (\ref{3.19}) taken with $n(R^+)\equiv 0$, it holds
\begin{equation} \label{3.27}
\frac{\partial}{\partial R} {\rm ln}\, \Xi_R(n,n_0=0) = 2\pi R n
- \frac{\kappa^2 R}{2} \sum_{l=0}^{\infty} \mu_l
\frac{K_{l-1}(\kappa R)}{I_{l-1}(\kappa R)} I_l^2(\kappa R) .
\end{equation}
The integration of this equation finally implies
\begin{eqnarray}
\fl {\rm ln}\, \Xi_R(n,n_0=0) = \pi R^2 n - \frac{1}{2}
{\rm ln}\left[ (\kappa R) I_1(\kappa R) \right] + {\rm const}
\nonumber \\ - \sum_{l=1}^{\infty} {\rm ln}
\left[ \left( \frac{2}{\kappa R} \right)^{l-1}
(l-1)! I_{l-1}(\kappa R) \right] \label{3.28} \\
+ \frac{1}{4} \sum_{l=0}^{\infty} \mu_l \left\{
\left[ (\kappa R)^2 +l^2 \right] I_l(\kappa R) K_l(\kappa R)
-(\kappa R)^2 I_l'(\kappa R) K_l'(\kappa R) - l \right\} . \nonumber
\end{eqnarray}
Here, since the grand partition function depends on the length
scale $a=1/\kappa_0$, the integration constant is in fact infinite
in the considered limit $a\to\infty$.
It is clear from the derivation procedure that formula (\ref{3.28})
is valid only for $R>0$, and it cannot serve as a basis for an
expansion around the $R=0$ point.

\subsection{Large-$R$ analysis}
Each of the above derived grand potentials
$\Omega = - (1/\beta) {\rm ln}\, \Xi$ is given in terms of infinite
sums which cannot be summed up explicitly.
What can be done is the evaluation of first few terms of the asymptotic
expansion of the sums for large disc radius $R\to\infty$.
Denoting by $\alpha$ either of the dimensionless combinations
$\kappa R$ or $\kappa_0 R$, one has to use the Debye expansion
\cite{Abramowitz} of the modified Bessel functions $I_l(\alpha), K_l(\alpha)$
and of their derivatives, since this expansion is valid for large $l$
{\em uniformly} with respect to $\alpha/l$.
In particular, one has
\numparts
\begin{eqnarray}
I_l(\alpha) = \frac{1}{\sqrt{2\pi}}
\frac{1}{(\alpha^2+l^2)^{1/4}}\, \rme^{\eta} \left[ 1+\frac{3t-5t^3}{24 l}
+ O\left( \frac{1}{\alpha^2+l^2} \right) \right] , \label{3.29a} \\
K_l(\alpha) = \sqrt{\frac{\pi}{2}}
\frac{1}{(\alpha^2+l^2)^{1/4}}\, \rme^{-\eta} \left[ 1-\frac{3t-5t^3}{24 l}
+ O\left( \frac{1}{\alpha^2+l^2} \right) \right] , \label{3.29b}
\end{eqnarray}
\endnumparts
and
\numparts
\begin{eqnarray}
I'_l(\alpha) = \frac{1}{\sqrt{2\pi}}
\frac{(\alpha^2+l^2)^{1/4}}{\alpha}\, \rme^{\eta}
\left[ 1+\frac{-9t+7t^3}{24 l}
+ O\left( \frac{1}{\alpha^2+l^2} \right) \right] , \label{3.30a} \\
K'_l(\alpha) = - \sqrt{\frac{\pi}{2}}
\frac{(\alpha^2+l^2)^{1/4}}{\alpha}\, \rme^{-\eta}
\left[ 1-\frac{-9t+7t^3}{24 l}
+ O\left( \frac{1}{\alpha^2+l^2} \right) \right] , \label{3.30b}
\end{eqnarray}
\endnumparts
where
$$
\eta(l,\alpha) = \sqrt{\alpha^2+l^2} - l\, {\rm sinh}^{-1}(l/\alpha),
\qquad t = l/\sqrt{\alpha^2+l^2} .
$$
The consequent sums over $l$ can be performed by applying the
(generalized) Euler-MacLaurin summation formula \cite{Abramowitz}:
\begin{equation} \label{3.31}
\fl \sum_{l=m}^{n} f(l) = \int_m^n f(l) \rmd l
+ \frac{1}{2}[ f(n)+f(m) ] + \frac{B_2}{2!} [ f'(n)-f'(m) ]
+ \frac{B_4}{4!}[ f'''(n)-f'''(m)] + \cdots ,
\end{equation}
where $B_{\nu}$ are Bernoulli numbers: $B_2=1/6$, $B_4=-1/30$, etc.

In the case of the Coulomb gas of particle density $n_0$, localized outside
of the disc of radius $R$ with vacuum in the disc hole, formula (\ref{3.24})
evaluated in the $R\to\infty$ limit implies
\begin{equation} \label{3.32}
\fl \beta \Omega_R(n=0,n_0) = - \beta p(n_0)(\vert \Lambda\vert -\pi R^2)
+ \beta \gamma(n_0)\, 2\pi R - \frac{1}{6} {\rm ln}(\kappa_0 R) + O(1) .
\end{equation}
The first term on the rhs of (\ref{3.32}) is the bulk contribution
with the pressure $p$ given by
\begin{equation} \label{3.33}
\beta p(n_0) = \left( 1-\frac{\beta}{4} \right) n_0 ,
\end{equation}
the second term is the surface contribution with the surface tension $\gamma$
given, in the Debye-H\"uckel limit, by \cite{Jancovici4}
\begin{equation} \label{3.34}
\beta \gamma(n_0) = \int_0^{\infty} \frac{\rmd l}{4\pi}
{\rm ln}\left[ \frac{(l+\sqrt{\kappa_0^2+l^2})^2}{4 l \sqrt{\kappa_0^2+l^2}}
\right] = \frac{\kappa_0}{8\pi} (4-\pi) .
\end{equation}
Finally, the third logarithmic term has the universal coefficient $-1/6$.
This Casimir term, caused by electrostatic fluctuations inside the wall,
tends to {\em dilate} the empty disc domain.
This is the fundamental difference in comparison with semi-infinite
geometries where fluctuating walls {\em attract} one another.

In the case of the Coulomb gas of particle density $n$, localized
inside the disc of radius $R$ and surrounded by vacuum,
formula (\ref{3.28}) evaluated in the $R\to\infty$ limit gives
\begin{equation} \label{3.35}
\fl \beta \Omega_R(n,n_0=0) = - \beta p(n)\, \pi R^2 +
\beta \gamma(n)\, 2\pi R + \frac{1}{6} {\rm ln}(\kappa R) + O(1) .
\end{equation}
As before, the first and second terms on the rhs  of (\ref{3.35})
correspond to the volume and surface parts of $\beta\Omega$, respectively.
The third logarithmic term has the coefficient $1/6$, in agreement
with the general relation (\ref{1.2}) taken at the disc value of $\chi=1$.
This coefficient is opposite to the one in (\ref{3.32}) which
confirms the relation of the universal term to the curvature
of the constraining surface.
It is important to note that in semi-infinite geometries (see e.g. paper I)
the grand potential of a Coulomb system constrained by vacuum plain hard
walls does not exhibit the universal Casimir term.
This is another fundamental difference between fully-finite and
semi-infinite geometries.

From a technical point of view, it is useful to sum the two expressions
for the grand potential (\ref{3.24}) and (\ref{3.28}), taken at
the same particle density denoted say by ${\bar n}$:
\begin{eqnarray}
\fl \beta \Omega_R(n={\bar n},n_0=0) + \beta \Omega_R(n=0,n_0={\bar n})
= \beta \Omega_{R=0} + {\rm const} \nonumber \\ + \frac{1}{2} {\rm ln}
\left[ ({\bar\kappa}R)^2 K_1({\bar\kappa}R) I_1({\bar\kappa}R) \right]
+ \sum_{l=1}^{\infty} {\rm ln} \left[ \frac{1}{2l}
({\bar\kappa} R)^2 I_{l-1}({\bar\kappa}R) K_{l+1}({\bar\kappa}R) \right] .
\label{3.36}
\end{eqnarray}
Using first the recursion formulas (3.10) and subsequently the
asymptotic expansions (3.29) and (3.30) for the modified Bessel functions,
the application of the Euler-MacLaurin summation formula (\ref{3.31})
implies after simple algebra that
\begin{equation} \label{3.37}
\fl \beta \Omega_R(n={\bar n},n_0=0) + \beta \Omega_R(n=0,n_0={\bar n})
= \beta \Omega_{R=0} + 2 \beta \gamma({\bar n}) 2 \pi R + O(1).
\end{equation}
This relation proves the consistency of asymptotic expansions
(\ref{3.32}) and (\ref{3.35}), and enables us to write down one
knowing the explicit form of the other.
Simplifying technicalities of this kind will be used in what follows.

We are now ready for studying the large-$R$ asymptotic behaviour of the
grand potential when both particle densities $n$ and $n_0$ are
nonzero, see formula (\ref{3.23}).
One has for the specific combination of grand potentials
\begin{eqnarray}
\fl \beta \Omega_R(n,n_0) -
\left[ \beta \Omega_R(n,n_0=0) + \beta \Omega_R(n=0,n_0) \right]
= {\rm const} + \frac{1}{2} {\rm ln} \left[
\frac{W_0}{\kappa \kappa_0 R I_1(\kappa R) K_1(\kappa_0 R)} \right]
\nonumber \\
- \sum_{l=1}^{\infty} {\rm ln}\left\{
\frac{[\kappa R I_{l-1}(\kappa R)] [\kappa_0 R K_{l+1}(\kappa_0 R)]}{2
l R W_l} \right\} . \label{3.38}
\end{eqnarray}
As above, using the recursion formulas and the asymptotic expansions
for the modified Bessel functions, the Euler-MacLaurin summation
formula leads to
\begin{eqnarray}
\fl \beta \Omega_R(n,n_0) -
\left[ \beta \Omega_R(n,n_0=0) + \beta \Omega_R(n=0,n_0) \right] \nonumber \\
\lo= R \int_0^{\infty} {\rm d}l\,
{\rm ln}\left[ \frac{2 l \left( \sqrt{\kappa^2+l^2}+\sqrt{\kappa_0^2+l^2}
\right)}{\left( l+\sqrt{\kappa^2+l^2}\right)\left(l+\sqrt{\kappa_0^2+l^2}
\right)} \right] + O(1) . \label{3.39}
\end{eqnarray}
With regard to the asymptotic expansions (\ref{3.32}) and (\ref{3.35}),
one finally arrives at
\begin{equation} \label{3.40}
\fl \beta \Omega_R(n,n_0) = -\beta p(n) \pi R^2 -
\beta p(n_0) ( \vert \Lambda\vert - \pi R^2) +
\beta \gamma(n,n_0) 2\pi R + O(1) ,
\end{equation}
where
\begin{equation} \label{3.41}
\beta \gamma(n,n_0) = \int_0^{\infty} \frac{\rmd l}{4\pi}
{\rm ln}\left[ \frac{\left( \sqrt{\kappa^2+l^2} + \sqrt{\kappa_0^2+l^2}
\right)^2}{4 \sqrt{\kappa^2+l^2} \sqrt{\kappa_0^2+l^2}} \right]
\end{equation}
is the (dimensionless) contact surface tension of the two 2D plasmas
in the Debye-H\"uckel limit, see equation (3.26) of paper I.
The universal logarithmic Casimir term does not appear in (\ref{3.40}),
it is ``screened''.
This phenomenon is intuitively expected: since there are Coulomb
systems on both sides of the disc boundary, the curvature contributions
with opposite signs cancel with one another.
The same cancellation of long-ranged Casimir forces takes place in
semi-infinite geometries, see paper I.

\section{The free-fermion point}

\subsection{General result}
The 2D Coulomb gas of symmetric unit charges is exactly solvable at
the collapse point $\beta = 2$; for general formalism see e.g.
section 4.1. of paper I.
In the grand canonical formalism, at $\beta=2$, the bulk system
is characterized by the rescaled particle fugacity $m=2\pi a z$
[$a$ is a free length scale introduced in (\ref{1.2})]
which has the dimension of an inverse length.
The many-particle densities can be expressed in terms of
specific Green functions $G_{qq'}({\bf r},{\bf r}')$ $(q,q'=\pm)$;
because of the symmetry between positive and negative particles
one only needs $G_{++}$ and $G_{-+}$.
These Green functions are determined by the equations
\begin{equation} \label{4.1}
( \Delta_1 - m^2 ) G_{++}({\bf r}_1,{\bf r}_2) =
- m \delta({\bf r}_1-{\bf r}_2)
\end{equation}
and
\begin{equation} \label{4.2}
G_{-+}({\bf r}_1,{\bf r}_2) = - \frac{1}{m}
\left( \frac{\partial}{\partial x_1} + \rmi \frac{\partial}{\partial y_1}
\right) G_{++}({\bf r}_1,{\bf r}_2) ,
\end{equation}
supplemented by the vanishing boundary conditions when
$\vert {\bf r}_1-{\bf r}_2 \vert \to \infty$.
In infinite space, the solution of (\ref{4.1}) reads
\begin{equation} \label{4.3}
G_{++}({\bf r}_1,{\bf r}_2) = \frac{m}{2\pi}
K_0(m\vert {\bf r}_1-{\bf r}_2 \vert) .
\end{equation}
The one-particle densities $n_+ = n_- = n/2$ ($n$ is the total
particle density), given by
\begin{equation} \label{4.4}
n_q({\bf r}) = m G_{qq}({\bf r},{\bf r}) ,
\end{equation}
are infinite since $K_0(m r)$ diverges logarithmically as $r\to 0$.
Regularization of the Coulomb interaction by a short-distance
cutoff $L$ implies for the particle density
\begin{equation} \label{4.5}
n = \frac{m^2}{\pi} K_0(m L)
{ \atop \stackrel{\sim}{\scriptstyle{m L \to 0}} }
\frac{m^2}{\pi} \left[ {\rm ln}\, \left( \frac{2}{m L} \right) - C \right] .
\end{equation}
In the inhomogeneous case when the system domain
$\Lambda = \cup_{\alpha} \Lambda^{(\alpha)}$, each subdomain
$\Lambda^{(\alpha)}$ is characterized by a constant rescaled fugacity,
$m({\bf r}) = m_{\alpha}$ for ${\bf r}\in \Lambda^{(\alpha)}$,
and the corresponding bulk density $n_{\alpha}$ defined as a function
of $m_{\alpha}$ by (\ref{4.5}).
Within each domain, the Green function $G_{++}$ obeys the
differential equation
\begin{equation} \label{4.6}
( \Delta_1 - m_{\alpha}^2 ) G_{++}({\bf r}_1,{\bf r}_2)
= - m_{\alpha} \delta({\bf r}_1-{\bf r}_2) , \qquad
{\bf r}_1 \in \Lambda^{(\alpha)} ,
\end{equation}
where the spatial position of the source point ${\bf r}_2$ is
arbitrary.
$G_{-+}$ is determined by relation (\ref{4.2}) with $m$ substituted
by the subdomain-dependent $m({\bf r}_1)$.
The boundary conditions are that both $G_{++}$ and $G_{-+}$ must be
continuous at each subdomain boundary $\partial \Lambda^{(\alpha)}$.
The one-particle densities are again given by (\ref{4.4})
with $m\to m({\bf r})$.

For the disc geometry of figure 1, the rescaled particle fugacity
is equal to $m$ in the disc domain $r<R$ and to $m_0$ in the
complementary wall $r>R$.
Let the source point ${\bf r}_2$ be first in the disc domain,
i.e. $r_2<R$.
Equations (\ref{4.1}) then take the form
\numparts
\begin{eqnarray}
( \Delta_1 - m^2 ) G_{++}({\bf r}_1,{\bf r}_2) =
- m \delta({\bf r}_1-{\bf r}_2) , \qquad & (r_1 < R) , \label{4.7a} \\
( \Delta_1 - m_0^2) G_{++}({\bf r}_1,{\bf r}_2) = 0 ,
& (r_1 > R) . \label{4.7b}
\end{eqnarray}
\endnumparts
In terms of polar coordinates, the solution of these equations
is written as an expansion of the form
\numparts
\begin{eqnarray}
\fl G_{++}({\bf r}_1,{\bf r}_2) = \frac{m}{2\pi} \sum_{l=-\infty}^{\infty}
\left[ I_l(m r_<) K_l(m r_>) \right. \nonumber \\
\left. + a_l I_l(m r_1) I_l(m r_2) \right]
\exp[{\rm i}l(\varphi_1-\varphi_2)], & ( r_1 < R), \label{4.8a} \\
\fl G_{++}({\bf r}_1,{\bf r}_2) = \frac{m}{2\pi} \sum_{l=-\infty}^{\infty}
b_l K_l(m_0 r_1) I_l(m r_2) \exp[{\rm i}l(\varphi_1-\varphi_2)], \qquad
& ( r_1 > R). \label{4.8b}
\end{eqnarray}
\endnumparts
The polar version of relation (\ref{4.2}) for $G_{-+}$ takes the form
\begin{equation} \label{4.9}
G_{-+}({\bf r}_1,{\bf r}_2) = - \frac{\rme^{\rmi \varphi_1}}{m({\bf r}_1)}
\left( \frac{\partial}{\partial r_1} + \frac{\rmi}{r_1}
\frac{\partial}{\partial\varphi_1} \right) G_{++}({\bf r}_1,{\bf r}_2) .
\end{equation}
The coefficients $a_l$ and $b_l$ are determined by the continuity
conditions for $G_{++}({\bf r}_1,{\bf r}_2)$ and
$G_{-+}({\bf r}_1,{\bf r}_2)$ at the disc boundary $r_1=R$:
\numparts
\begin{eqnarray}
K_l(mR) + a_l I_l(mR) = b_l K_l(m_0 R) , \label{4.10a} \\
K_{l+1}(mR) - a_l I_{l+1}(mR) = b_l K_{l+1}(m_0 R) . \label{4.10b}
\end{eqnarray}
\endnumparts
The solution reads
\numparts
\begin{eqnarray}
a_l = - \frac{K_l(mR)}{I_l(mR)} + \frac{1}{m R V_l}
\frac{K_l(m_0 R)}{I_l(mR)} , \label{4.11a} \\
b_l = \frac{1}{m R V_l} ,
\end{eqnarray}
\endnumparts
where
\begin{equation} \label{4.12}
V_l = I_l(mR) K_{l+1}(m_0 R) + I_{l+1}(m R) K_l(m_0 R) .
\end{equation}
Note the symmetry $V_l = V_{-l-1}$.
It can be readily shown that the auxiliary quantity $V_l$ fulfills
the equation
\begin{equation} \label{4.13}
\frac{\partial}{\partial R} {\rm ln}\, ( R V_l) = \frac{m-m_0}{V_l}
\left[ I_l(m R) K_l(m_0 R) + I_{l+1}(m R) K_{l+1}(m_0 R) \right] .
\end{equation}
One proceeds analogously when the source point ${\bf r}_2$ lies
in the wall, i.e. $r_2 > R$.
For the case $r_1 > R$, one gets
\begin{eqnarray}
\fl G_{++}({\bf r}_1,{\bf r}_2) = \frac{m_0}{2\pi} \sum_{l=-\infty}^{\infty}
\left[ I_l(m_0 r_<) K_l(m_0 r_>) \right. \nonumber \\
\left. + c_l K_l(m_0 r_1) K_l(m_0 r_2) \right]
\exp[\rmi l(\varphi_1-\varphi_2)], \qquad ( r_1, r_2 > R) . \label{4.14}
\end{eqnarray}
The coefficients $c_l$ are given by the boundary conditions at
$r_1=R$ as follows
\begin{equation} \label{4.15}
c_l = - \frac{I_l(m_0 R)}{K_l(m_0 R)} + \frac{1}{m_0 R V_l}
\frac{I_l(m R)}{K_l(m_0 R)} .
\end{equation}

The density at $r\to R^-$, determined by (\ref{4.4}), (\ref{4.8a})
and (\ref{4.11a}), reads
\begin{equation} \label{4.16}
n(R^-) = \frac{m}{\pi R} \sum_{l=-\infty}^{\infty}
\frac{1}{V_l} I_l(m R) K_l(m_0 R) .
\end{equation}
The sum in (\ref{4.16}) is divergent, and so it has to be formally
regularized by taking an upper cutoff on $l$.
Similarly, the density at $r\to R^+$, determined by (\ref{4.4}),
(\ref{4.14}) and (\ref{4.15}), is expressible as follows
\begin{equation} \label{4.17}
n(R^+) = \frac{m_0}{m} n(R^-) .
\end{equation}
The grand canonical partition function $\Xi_R(m,m_0)$ is again
determined by the differential equation (\ref{3.19}), in particular
\begin{equation} \label{4.18}
\fl \frac{\partial}{\partial R} {\rm ln}\, \Xi_R =
2 (m - m_0) \sum_{l=0}^{\infty} \frac{1}{V_l}
\left[ I_l(m R) K_l(m_0 R) + I_{l+1}(m R) K_{l+1}(m_0 R) \right] .
\end{equation}
Due to the equality (\ref{4.13}),
this equation can be integrated to the form
\begin{equation} \label{4.19}
{\rm ln}\, \Xi_R(m,m_0) = {\rm const} +
2 \sum_{l=0}^{\infty} {\rm ln}(R V_l) .
\end{equation}
The integration constant is fixed by the $R\to 0$ limit.
Since $\lim_{R\to 0} R V_l = m_0^{-1} (m/m_0)^l$ for $l\ge 0$,
one gets
\begin{eqnarray}
\fl {\rm ln}\, \Xi_R(m,m_0) = {\rm ln}\, \Xi_{R=0} +
2 \sum_{l=0}^{\infty} {\rm ln}\Bigg\{ \left( \frac{m_0}{m} \right)^l
m_0 R  \nonumber \\
\times \left[ I_l(m R) K_{l+1}(m_0 R) + I_{l+1}(m R) K_l(m_0 R) \right]
\Bigg\} . \label{4.20}
\end{eqnarray}
Here, $\Xi_{R=0}$ is the grand canonical partition function of
the system with zero disc radius, i.e.
$\lim_{\Lambda\to\infty} {\rm ln}\, \Xi_{R=0}/\vert \Lambda\vert
= \beta p(n_0)$.

Although (\ref{4.20}) was derived under the assumption
that both rescaled fugacities $m$ and $m_0$ are nonzero, there is
no problem to consider the zero limit of either of fugacities
directly in (\ref{4.20}).
One obtains
\begin{equation} \label{4.21}
{\rm ln}\, \Xi_R(m,m_0=0) = 2 \sum_{l=0}^{\infty} {\rm ln}
\left[ l! \left( \frac{2}{mR}\right)^l I_l(m R) \right]
\end{equation}
and
\begin{equation} \label{4.22}
{\rm ln}\, \Xi_R(m=0,m_0) = {\rm ln} \Xi_{R=0}
+ 2 \sum_{l=0}^{\infty} {\rm ln}\left[
\frac{2}{l!} \left( \frac{m_0 R}{2} \right)^{l+1}
K_{l+1}(m_0 R) \right] .
\end{equation}

\subsection{Large-$R$ analysis}
The result (\ref{4.21}) for the 2D two-component plasma at
$\beta=2$, in the disc surrounded by vacuum, has already been
derived in \cite{Jancovici1}.
There, the large-$R$ asymptotic form of the grand potential
was derived in the form
\begin{equation} \label{4.23}
\beta \Omega_R(m,m_0=0) = -\beta p(m) \pi R^2 + \beta \gamma(m) 2\pi R
+ \frac{1}{6} {\rm ln}( m R ) + O(1) ,
\end{equation}
where $p(m)$ is the regularized pressure and the surface tension
$\gamma(m)$ is given by
\begin{equation} \label{4.24}
\beta \gamma(m) = \int_0^{\infty} \frac{\rmd l}{2\pi}
{\rm ln}\left( \frac{2\sqrt{m^2+l^2}}{l+\sqrt{m^2+l^2}} \right)
= m \left( \frac{1}{4} - \frac{1}{2\pi} \right) .
\end{equation}

One can apply the standard procedure to formula (\ref{4.22})
corresponding to the plasma outside of the empty disc, to obtain
the large-$R$ asymptotic behaviour
\begin{equation} \label{4.25}
\fl \beta \Omega_R(m=0,m_0) = -\beta p(m_0)(\vert \Lambda\vert - \pi R^2)
+ \beta \gamma(m_0) 2\pi R - \frac{1}{6} {\rm ln}( m_0 R ) + O(1) .
\end{equation}
One sees that, similarly as in the Debye-H\"uckel limit, in comparison
with (\ref{4.23}) the universal logarithmic term has the opposite sign.

Technically, it is simpler to derive (\ref{4.25}) by summing up
the two basic expressions (\ref{4.21}) and (\ref{4.22}) taken
at the same rescaled fugacity, say ${\bar m}$:
\begin{eqnarray}
\fl \beta \Omega_R(m={\bar m},m_0=0) + \beta \Omega_R(m=0,m_0={\bar m})
\nonumber \\ \lo= \beta \Omega_{R=0} - 2 \sum_{l=0}^{\infty}
{\rm ln}\left[ {\bar m} R I_l({\bar m}R) K_{l+1}({\bar m}R) \right] .
\label{4.26}
\end{eqnarray}
The standard asymptotic procedure for the modified Bessel functions
leads to the following asymptotic behaviour
\begin{equation} \label{4.27}
\fl \beta \Omega_R(m={\bar m},m_0=0) + \beta \Omega_R(m=0,m_0={\bar m})
= \beta \Omega_{R=0} + 2 \beta \gamma({\bar m}) 2\pi R + O(1) ,
\end{equation}
which proves the consistency of relations (\ref{4.23}) and (\ref{4.25}).

Finally, considering the special combination of grand potentials
\begin{eqnarray}
\fl \beta \Omega_R(m,m_0) - \left[ \beta \Omega_R(m,m_0=0)
+ \beta \Omega_R(m=0,m_0) \right] \nonumber \\
\lo= - 2 \sum_{l=0}^{\infty} {\rm ln} \left[
1 + \frac{I_{l+1}(m R) K_l(m_0 R)}{I_l(m R) K_{l+1}(m_0 R)} \right] ,
\label{4.28}
\end{eqnarray}
and using the standard asymptotic procedure, one gets
\begin{eqnarray}
\fl \beta \Omega_R(m,m_0) - \left[ \beta \Omega_R(m,m_0=0)
+ \beta \Omega_R(m=0,m_0) \right] \nonumber \\
\lo= R \int_0^{\infty} \rmd l\, {\rm ln}\left[
\frac{\left( l+\sqrt{m^2+l^2}\right)\left( l+\sqrt{m_0^2+l^2}\right)}{2
\left( m m_0 + \sqrt{m^2+l^2} \sqrt{m_0^2+l^2} +l^2\right)} \right] + O(1) .
\label{4.29}
\end{eqnarray}
With regard to the asymptotic relations (\ref{4.23}) and (\ref{4.25}),
one concludes that
\begin{equation} \label{4.30}
\fl \beta \Omega_R(m,m_0) = - \beta p(m) \pi R^2 - \beta p(m_0)
(\vert \Lambda\vert - \pi R^2) + \beta \gamma(m,m_0) 2\pi R + O(1) ,
\end{equation}
where
\begin{equation} \label{4.31}
\beta \gamma(m,m_0) = \int_0^{\infty} \frac{\rmd l}{2\pi}
{\rm ln}\left( \frac{2 \sqrt{m^2+l^2} \sqrt{m_0^2+l^2}}{m m_0
+ \sqrt{m^2+l^2} \sqrt{m_0^2+l^2} + l^2}\right)
\end{equation}
defines the surface tension of the two 2D plasmas at $\beta=2$,
see equation (4.12) of paper I.
As was expected, the universal logarithmic term disappears once again.

We have verified that the basic features of the mean-field behaviour,
predicted by the Debye-H\"uckel analysis in the previous section,
persist also at the specific inverse temperature $\beta=2$.

\section{Conclusion}
The name ``ideal-conductor'' usually means that the electric potential has
some constant value (for instance zero) inside the conductor, without any
fluctuations. 
More precisely, we call such a model of a conductor 
``inert ideal-conductor''. 
The present paper deals with microscopic models of conductors, in
which there are fluctuations of the electric potential. 
They are ``living'' conductors. 
In the limit when the microscopic lengths, such as the Debye length, 
go to zero (in practice, are small compared to the macroscopic length, 
here such as the radius $R$), the fluctuations of the electric
potential survive. 
We call a conductor in that high-density limit ``living ideal-conductor''.

In a previous publication \cite{Jancovici2}, Coulomb systems with
inert ideal-conductor boundary conditions were studied\footnote{In 
\cite{Jancovici2}, the Coulomb potential in $\nu$ dimensions $(\nu>2)$ 
was defined as $r^{2-\nu}$ while in I it was defined as $r^{2-\nu}/(\nu-2)$.
These different definitions do not change the Casimir terms under
consideration.}.
In this model, for a Coulomb system in a slab of width $d$, a
repulsive Casimir force was found, which is just the opposite of 
(\ref{2.8}) which holds for living ideal-conductor plates
separated by vacuum. 
For a Coulomb system in a slab with living ideal-conductor walls, 
the two contributions cancel each other, leaving only a
short-range attraction, as shown in I.

Similarly, in ref. \cite{Jancovici2}, a 2D Coulomb system in a disc 
of radius $R$ with inert ideal-conductor walls was found to have 
a logarithmic universal contribution $(1/6)\ln R$ to $\beta$ times 
its grand potential. 
This is just the opposite of the $-(1/6)\ln R$ found here for an empty 
circular hole surrounded by a Coulomb gas. 
Again, for a Coulomb system in a disc with living
conductor walls, the two contributions cancel each other.

\ack
The authors acknowledge support from the CNRS-SAS agreement,
Project No. 14439.
A partial support of L. {\v S}amaj by a VEGA grant is acknowledged.

\appendix
\setcounter{section}{1}
\section*{Appendix}

In ref. \cite{Jancovici2}, (\ref{2.8}) has been derived
by using the quantum Hamiltonian in $\nu-1$ dimensions.
Here, we give an alternative, more direct, derivation.

The free energy per unit area $F$ corresponding to the Gaussian
partition function (\ref{2.7}), generalized to $\nu$ dimensions,
is given, up to an irrelevant additive constant, by
\begin{equation} \label{A.1}
\beta F=-\frac{1}{L^{\nu-1}}\ln Z_{\rm G}=
\frac{1}{2L^{\nu-1}}\sum\ln\lambda,
\end{equation}
where $L$ is the linear size of a plate and $\lambda$ the eigenvalues
of minus the Laplacian.
The planes, on which $\phi$ obeys Dirichlet boundary conditions,
are perpendicular to the $x$ axis at $x=0$ and $x=d$.
Each point is defined by its Cartesian coordinates
$(x,{\bf r}^{\perp})$ where ${\bf r}^{\perp}$ is a
$(\nu-1)$-dimensional vector normal to the $x$ axis.
In this geometry, in the limit of $L$ infinite, the eigenfunctions
of $-\Delta$ are
$\exp(-{\rm i}{\bf l}\cdot{\bf r}^{\perp})\sin(n\pi x/d)$
where ${\bf l}$ is a wavevector normal to the $x$ axis and $n$ an
integer larger than 0.
The corresponding eigenvalues are $\lambda=l^2+(n\pi/d)^2$.
Therefore, (\ref{A.1}) can be written as
\begin{equation} \label{A.2}
\beta F=\frac{1}{2}\int\frac{{\rm d}^{\nu-1}l}{(2\pi)^{\nu-1}}
\sum_{n=1}^{\infty}\ln\left(l^2+\left( \frac{n\pi}{d} \right)^2 \right).
\end{equation}
Thus
\begin{equation} \label{A.3}
-\beta\frac{\partial F}{\partial d}=
\int\frac{{\rm d}^{\nu-1}l}{(2\pi)^{\nu-1}}\frac{1}{d}
\sum_{n=1}^{\infty}\frac{(n\pi /d)^2}{l^2+(n\pi /d)^2}.
\end{equation}

The sum on $n$ in (\ref{A.3}) diverges.
The divergent part can be separated by noting that the summand minus 1 is
$-(ld/\pi)^2/[(ld/\pi)^2+n^2]$, which can be explicitly summed
\cite{Gradshteyn}.
The result for the total sum, including the divergent part, is
\begin{equation} \label{A.4}
\frac{1}{d}\sum_{n=1}^{\infty}\frac{(n\pi /d)^2}{l^2+(n\pi /d)^2}=
-\frac{1}{2}l\coth(ld)+\frac{1}{2d}+\sum_{n=1}^{\infty}\frac{1}{d}.
\end{equation}
The last two terms of (\ref{A.4}) can be regrouped into the divergent
sum $S = (1/2)\sum_{n=-\infty}^{\infty}(1/d)$.
Since $F$ has a bulk term of the form $Ad$, with $A$ infinite in
the present ideal conductor limit, for obtaining the finite-$d$
Casimir force $f$, we must subtract from (\ref{A.3}) its value for
$d$ infinite.
For $d$ infinite, the term $n\pi/d$ in the eigenvalue $\lambda$ becomes
the continuous wave number $k$ and the divergent sum $S$ becomes
$\int_{-\infty}^{\infty}{\rm d}k/(2\pi)$.
Since, for $d$ finite, the summand in $S$ is a constant,
$S$ can also be written as the same integral on $k$, and after the
subtraction the divergent term disappears.
Finally,
\begin{eqnarray}
\beta f&=&-\left[\beta\frac{\partial F}{\partial d}
-\left.\beta\frac{\partial F}{\partial d}\right|_{d=\infty}\right]
\nonumber \\
&=&-\frac{1}{2}\int\frac{{\rm d}^{\nu-1}l}{(2\pi)^{\nu-1}}l[\coth(ld)-1]
=-\frac{s_{\nu-1}\Gamma(\nu)\zeta(\nu)}{(2\pi)^{\nu-1}(2d)^{\nu}},
\label{A.5}
\end{eqnarray}
in agreement with (\ref{2.8}).


\section*{References}

\begin{thebibliography}{33}

\bibitem{Duplantier1} Duplantier B, 2003
{\it Poincar\'e Seminar 2002} ed B Duplantier and V Rivasseau
(Basel: Birkh\"auser) p 53

\bibitem{Lifshitz} Lifshitz E M, 1954
{\it Sov. Phys. JETP} {\bf 2} 73
\nonum
Dzyaloshinskii I E, Lifshitz E M and Pitaevskii L P, 1961
{\it Sov. Phys. Usp.} {\bf 73} 381

\bibitem{Schwinger1} Schwinger J, DeRaad L L and Milton K A, 1978
{\it Ann. Phys., NY} {\bf 115} 1

\bibitem{Balian1} Balian R and Duplantier B, 1977
{\it Ann. Phys., NY} {\bf 104} 300
\nonum
Balian R and Duplantier B, 1978
{\it Ann. Phys., NY} {\bf 112} 165

\bibitem{Mostepanko} Mostepanko V M and Trunov N N, 1997
{\it The Casimir Effect and Its Application}
(Oxford: Clarendon)

\bibitem{Bordag} Bordag M, Mohideen U and Mostepanko V M,
{\it Phys. Rep.} {\bf 353} 1

\bibitem{Feinberg} Feinberg J, Mann A and Revzen M, 2001
{\it Ann. Phys., NY} {\bf 288} 103

\bibitem{Schwinger2} Schwinger J, 1975
{\it Lett. Math. Phys.} {\bf 1} 43

\bibitem{Balian2} Balian R, 2003
{\it Poincar\'e Seminar 2002} ed B Duplantier and V Rivasseau
(Basel: Birkh\"auser) p 71

\bibitem{Blote} Bl\"ote H W J, Cardy J L and Nightingale M P, 1986
{\it Phys. Rev. Lett.} {\bf 56} 742

\bibitem{Affleck} Affleck I, 1986
{\it Phys. Rev. Lett.} {\bf 56} 746

\bibitem{Duplantier2} Duplantier B and David F, 1988
{\it J. Stat. Phys.} {\bf 51} 327

\bibitem{Cardy} Cardy J L and Peschel I, 1988
{\it Nucl. Phys.} B {\bf 300} 377

\bibitem{Houches} Cardy J L, 1990 in {\it Fields, Strings and Critical
Phenomena, Les Houches 1988}, Br\'ezin E and Zinn-Justin J eds.
(Amsterdam: North-Holland)

\bibitem{Forrester} Forrester P J, 1991
{\it J. Stat. Phys.} {\bf 63} 491

\bibitem{Jancovici1} Jancovici B, Manificat G and Pisani C, 1994
{\it J. Stat. Phys.} {\bf 76} 307

\bibitem{Jancovici2} Jancovici B and T\'ellez G, 1996
{\it J. Stat. Phys.} {\bf 82} 609

\bibitem{Tellez} T\'ellez G, 2001
{\it J. Stat. Phys.} {\bf 104} 945

\bibitem{Torres} Torres A and T\'ellez G, 2004
{\it J. Phys. A: Math. Gen.} {\bf 37} 2121
\nonum
Torres A and T\'ellez G, 2004
cond-mat/0404588, to appear in {\it J. Stat. Phys.}

\bibitem{Cornu} Cornu F and Jancovici B, 1987
{\it J. Stat. Phys.} {\bf 49} 33
\nonum
Cornu F and Jancovici B, 1989
{\it J. Chem. Phys.} {\bf 90} 2444

\bibitem{Jancovici3} Jancovici B, 2000
{\it J. Stat. Phys.} {\bf 100} 201

\bibitem{Samaj1} {\v S}amaj L and Jancovici B, 2002
{\it J. Stat. Phys.} {\bf 106} 323

\bibitem{Janco} Jancovici B, Kalinay P and {\v S}amaj L, 2000
{\it Physica} A {\bf 279} 260

\bibitem{Buenzli} Buenzli P R and Martin Ph A, 2004
cond-mat/0407808, to appear in {\it J. Stat. Phys.}

\bibitem{Attard} Attard P, Mitchell D J and Ninham B W, 1988
{\it J. Chem. Phys.} {\bf 88} 4987
\nonum
Attard P, Mitchell D J and Ninham B W, 1988
{\it J. Chem. Phys.} {\bf 89} 4358

\bibitem{Dean} Dean D S and Horgan R R, 2003
{\it Phys. Rev.} E {\bf 68} 051104
\nonum
Dean D S and Horgan R R, 2003
{\it Phys. Rev.} E {\bf 68} 061106

\bibitem{Jancovici4} Jancovici B and {\v S}amaj L, 2004
{\it J. Stat. Mech.} P08006, here referred to as paper I

\bibitem{Samuel} Samuel S, 1978 {\it Phys. Rev.} D {\bf 18} 1916

\bibitem{Minnhagen} Minnhagen P, 1987 {\it Rev. Mod. Phys.} {\bf 59} 1001

\bibitem{Samaj2} {\v S}amaj L, 2003 {\it J. Phys.} A {\bf 36} 5913

\bibitem{Zamolodchikov} A. B. Zamolodchikov and Al. B. Zamolodchikov, 1979
{\it Ann. Phys., NY} {\bf 120} 253

\bibitem{Caillol} Caillol J M, 2004 {\it J.Stat.Phys.} {\bf 115} 1461

\bibitem{Gradshteyn} Gradshteyn I S and Ryzhik I M, 1994
{\it Table of Integrals, Series and Products} 5th edn.
(London: Academic Press)

\bibitem{Erdelyi} Erd\'elyi A, 1953
{\it Higher Transcendental Functions} vol 2 (New York: McGraw-Hill)

\bibitem{Jancovici5} Jancovici B, 2003
{\it J. Stat. Phys.} {\bf 110} 879, Appendix B

\bibitem{Abramowitz} Abramowitz M and Stegun I A, 1964
{\it Handbook of Mathematical Functions} (Washington D.C.:
National Bureau of Standards)

\end{thebibliography}
\end{document}